\documentclass[12pt,letterpaper]{article}

\usepackage{amsmath,amssymb,calc}
\usepackage{graphicx,color}

\newcommand\be{\begin{equation}}
\newcommand\ee{\end{equation}}
\newcommand{\bea}{\begin{eqnarray}}
\newcommand{\eea}{\end{eqnarray}}
\newcommand{\half}{\frac{1}{2}}

\newcommand{\alp}{\alpha' }

\newcommand{\nn}{\nonumber}

\newcommand{\dbar}{{\overline{\rm D}}}
\newcommand{\ddbpm}{{\rm D}_{p-1}-{\dbar}_{p-1}    }
\newcommand{\ddbp}{ {\rm D}_p-{\dbar}_p     }

\newcommand{\ddbn}{ {\rm D}_9-{\dbar}_9     }
\newcommand{\ddbe}{ {\rm D}_8-{\dbar}_8     }

\def\id{\protect{{1 \kern-.28em {\rm l}}}}

\def\unit{\relax{\rm 1\kern-.26em I}}

\def\id{\protect{{1 \kern-.28em {\rm l}}}}

\begin{document}
\begin{titlepage}

\begin{center}
\hfill QMUL-PH-08-04 \\
\vskip 15mm

{\Large {\bf  Phase Transitions in Separated ${\rm D}_{p-1}$ and anti-${\rm D}_{p-1}$ \,Branes at Finite Temperature}}\\

\vskip 10mm

{\bf Vincenzo Cal\`o }
and {\bf Steven Thomas}

\vskip 4mm
{\em Center for Research in String Theory}\\
{\em Department of Physics, Queen Mary, University of London}\\
{\em Mile End Road, London, E1 4NS, UK.}\\
{\tt v.calo@qmul.ac.uk, s.thomas@qmul.ac.uk}\\

\vskip 6mm

\end{center}

\vskip .1in

\begin{center} {\bf Abstract }\end{center}

\begin{quotation}\noindent

We consider a pair of parallel ${\rm D}_{p-1}$ and anti-${\rm D}_{p-1}$ branes in flat space, with a finite separation $d$ along some perpendicular spatial direction and at finite temperature.  If this spatial direction  is compactified on a circle
then by T-duality, the system is equivalent to a ${\rm D}_{p}$-anti\,${\rm D}_{p} $ pair wrapped around the dual circle  with a constant Wilson line $A \approx  d $ on one of the branes.
We focus in particular on the $p=9$ case and compute the free energy of this system and study the occurrence of second order phase transitions as both the temperature and Wilson line (brane-antibrane separation) are varied. 
In the limit of vanishing Wilson line we recover the previous results obtained in the literature, whereby the open string vacuum at the origin of the tachyon field $T=0$
is stabilized at sufficiently high temperature at which a second order phase transition occurs.
For sufficiently large Wilson line, we find new second order phase transitions corresponding to the existence of two minima in the tachyon effective potential at finite temperature and tachyon field value. Entropic arguments suggest that as the system cools, the tachyon is likely to find itself in the minimum that approaches infinity as the temperature vanishes (i.e. the one corresponding to the closed string vacuum), rather than the 
minimum at $T=0$ (corresponding to the open string vacuum).
\end{quotation}
\end{titlepage}
\eject

\tableofcontents

\section{Introduction}
\setcounter{equation}{0}
The study of unstable (non-BPS) D-brane configurations in flat space \cite{sen} has been 
a fertile area of research in recent years. 
Sen's conjectures \cite{sen} concerning what happens to unstable D-branes and the fate
 of the open string vacuum has been supported by results in boundary string field theory, 
 (BSFT) \cite{Witten}, \cite{bsft}, \cite{schnabl}. 
 In this picture a coincident parallel brane-antibrane configuration is unstable to decay
  through the open string tachyon field $T$ rolling down to its minimum and thus producing enough negative energy to cancel that coming from the brane tensions. Thus the open string vacuum decays to that of the closed string. 
A comprehensive review of tachyon condensation from the point of view of BSFT and other approaches can be found in \cite{sen2}.

Whilst the above behaviour of the open string tachyon is true for a system at zero temperature there have been several papers discussing the situation if the brane-antibrane pair is considered as part of a thermodynamic system at finite temperature \cite{Danielsson}, \cite{Hotta1}, \cite{Hotta2}, \cite{Hashimoto}, \cite{Huang}. Including finite temperature effects is interesting because 
 it's possible that such  brane configurations could survive in the early universe and thus be stable at finite temperature \cite{Hotta3}.
 
In \cite{Hotta1},  Hotta investigated the phase structure of a finite temperature ${\rm D}_p$\,brane-anti\,${\rm D}_p$ brane pair (which we will abbreviate in this paper as 
${\rm D}_p-{\dbar}_p $) where the branes were assumed to be coincident and in flat space. Using the framework of boundary string field theory (BSFT) he showed that in the  $p=9$ case, a phase transition occurs just below the string Hagedorn temperature, whereas for $p<9 $ there is no phase transition. In the $p=9$ case, the zero temperature minimum of the tachyon effective potential  was shown to shift from $T \rightarrow \infty$  
towards $T \rightarrow 0 $ as the temperature approached criticality. Thus the interpretation is that the open string vacuum is stabilized at sufficiently high temperature (but below the Hagedorn transition) in the case of $\ddbn $ whereas for pairs of lower dimensionality, no such transition occurs and the point $T=0$ remains unstable at high temperature.

These results are in broad agreement with those of Danielsson et al \cite{Danielsson} who investigated the same system but rather than including the full set of string states, they focussed on the truncation to the tachyonic sector only, in computing the free energy. 

In this paper we wish to generalize the results above to the case where the $\ddbpm$ pair is separated along some perpendicular spatial direction, but still parallel and in flat space. We shall assume that 
the pair has a finite separation $d$ along a perpendicular spatial direction  which is compactified on a circle $S^1$. Then by T-duality, the system is equivalent to a $\ddbp$ pair wrapped around the dual circle $\tilde{S^1} $, with a constant Wilson line $A \approx d $ turned on one of the branes \cite {Hashimoto}, \cite{sen3}.

At zero temperature, one may extend the BSFT results to include Wilson lines and obtain an expression for the 
effective potential at 1-loop $V_{eff} (T, A)$ depending on $T$ and the Wilson line $A$. 
At tree level, the extrema of this potential depend on the size of $A$. For $ A < \frac{1}{\sqrt{2 \alp}} $ the potential has a local maximum at $T=0$ as in the case of a coincident brane-antibrane. If $A > \frac{1}{\sqrt{2 \alp}} $, $T=0$ becomes a local minimum and so the open string vacuum is metastable \cite{Hashimoto}.  We shall see that when we consider this latter situation at finite temperature, we have an interesting situation whereby  the effective potential (in the canonical ensemble) has 
two local minima at finite values of $T$, which then approach the values $T=0, \infty$, as the temperature approaches zero. Thus we can ask the question what vacuum the tachyon field will likely be found, i.e., which vacuum is 
thermodynamically favoured over the other? Also what is the likelihood of  a first order phase transition occurring since we anticipate that the two minima might well become degenerate at some particular temperature, so that quantum tunnelling may become important. 

The structure of this paper is as follows. In section 2 we review the 
two derivative truncation of the BSFT approach to studying the tachyon 
potential for coincident $\ddbpm$ system \cite{Witten}, \cite{bsft} and  its extension 
to the case of finite separation (or addition of Wilson lines being included in the world-volume action of the T-dual wrapped $\ddbp$ 
system).
In section 3 we consider the 1-loop (annulus or cylinder) computations of 
the free energy of open strings stretched between separated $\ddbpm$
pair. In section 4 we then investigate the critical points of the free energy
and determine the nature of the phase transitions as the temperature approaches the
 Hagedorn temperature from below.  In particular we compare the situation when 
 the Wilson line modulus $A$ is greater than or less than its critical value
 $A_{crit} = \frac{1}{\sqrt{2 \alp}}$ . Finally in section 5 we draw some conclusions 
 from our results.
 
\section{Two-derivative truncation of the BSFT of the $D\bar{D}$ system}
\label{SecTwo}
\setcounter{equation}{0}

In string theory a pair of parallel $\ddbp$ pair constitutes an unstable object.\\
To study the dynamics of unstable D-branes, the BSFT \cite{Witten}, \cite{bsft} is a useful tool and it has provided a good understanding of tachyon condensation at the classical level. It describes the off-shell dynamics of open strings in a fixed on-shell background of closed strings in which an open string field configuration corresponds to a boundary term in the world-sheet action of the string. Therefore, specifying a boundary term means giving the background values of the various modes of the open string. It is based on the Batalin-Vilkovisky formalism whose master equation provides the effective action of the theory.
In the bosonic string theory, the disk partition function of the open string theory $Z$ and the BSFT action are related by the master equation
\be
S=\left(1+ \beta^i \frac{\partial}{\partial g^i}  \right) Z
\label{BosMastEq}
\ee
where $g^i$ are the couplings of the boundary interactions and $\beta^i$ are the corresponding world-sheet $\beta$-functions.
Given a specific form of the tachyon profile, the BSFT action reduces to the effective action for the tachyon field allowing us to compute the tree level tachyon potential. For superstrings, the tachyon $\beta$-function is zero and eq. (\ref{BosMastEq}) reduces to
\be
S = Z
\label{BSFTandAction}
\ee
The partition function $Z$ was computed in \cite{Kraus}. Let us briefly review their results here. The disc partition function is formally defined as
\be
Z= \int\, {\cal D}X {\cal D}\psi \, e^{-\left( S_{\textrm{bulk}}+S_{\textrm{bndy}}\right)} 
\label{BasicZ}
\ee
where
\be
S_{\textrm{bulk}} = \frac{1}{4 \pi} \int \, d^2z\, \left( \frac{2}{ \alpha^{\prime} } \partial X^{\mu} \bar{\partial}X_{\mu} + \psi^{\mu} \bar{\partial} \psi_{\mu} + \bar{\psi}^{\mu} \partial \bar{\psi}_{\mu}  \right)
\label{BulkAction}
\ee
is the bulk action for the NSR string.

The boundary term of the $\ddbp$ system is computed introducing auxiliary boundary fermion superfields $\Gamma^I = \eta^I +\theta F^I$ where $I=1,\,\,2^m$, and, $N=2^{m-1}$ is the number of $\ddbp$ pairs. Consider, for example, the case where we have $2^m$ branes. The $2^m\times 2^m$ matrices of the gauge group $U(2^m)$, generated by the branes, can be expanded in terms of $SO(2m)$ gamma matrices. Now, instead of gamma matrices, one can introduce $2m$ boundary fermion superfields $\Gamma^I$ with action $S=-\int d\tau d\theta \frac{1}{4} \Gamma^I D \Gamma^I$, and, after canonically quantizing, one arrives at the anti-commutation relations $\{ \eta^I ,\eta^J \} = 2 \delta^{IJ}$. Thus, $\eta^I$ can represent the Clifford algebra needed for the expansion of the $2^m \times 2^m$ matrices \cite{Kraus}.\\ In the case, e.g., of a single $\ddbn$ pair, expanding the resulting action in terms of the component fields one has
\bea
S_{\textrm{bndy}} &=& - \int \, \left[ -\frac{\alpha^{\prime}}{4} T^I T^I + \frac{1}{4} \dot{\eta}^I \eta^I + \frac{\alpha^{\prime}}{2} D_{\mu}  T^I \psi^{\mu} \eta^I +\frac{i}{2} \left( \dot{X}^{\mu} A_{\mu} + \half F_{\mu \nu} \psi^{\mu} \psi^{\nu} \right)   \right.  \nonumber \\ &\phantom{1}& \left. \phantom{123456} +\frac{i}{4} \left( \dot{X}^{\mu} A_{\mu}^{IJ} +\half \alpha^{\prime} F_{\mu \nu}^{IJ} \psi^{\mu} \psi^{\nu} \right) \eta^{I}\eta^{J} \right] d\tau
\label{BdryAction}
\eea
Here $I,J =1,2$,
\bea
A_{\mu}^{\pm} &=& \half \left(A_\mu \pm i A_{\mu}^{12} \right) \nn \\
D_{\mu} T^{I} &=& \partial_{\mu} T^I - iA_{\mu}^{IJ}T^J 
\eea
and the gauge fields $A_{\mu}^{\pm}$ on the brane and anti-brane, respectively, have been expressed in terms of the abelian gauge fields $A_{\mu}^{IJ}$, (anti-symmetrized in $I,J$) and $A_{\mu}$. Moreover, $F_{\mu \nu} = \partial_{[\mu} A_{\nu ]} $ and 
$ {F^{IJ}}_{\mu \nu} = \partial_{[\mu} {A^{IJ}}_{\nu ]} $. 

In the case of a constant tachyon field and zero gauge fields, the boundary action reduces to
\be
S_{\textrm{bdry}}= \frac{\alpha^{\prime}}{4} \int \, d \tau \,  T^I T^I
\ee
Since there is no other dependence on the tachyon field in the bulk action, we learn that in this case the tachyon potential for the $\ddbn$ system is
\be
V_0(T)=2 T_9 \, e^{-2 \pi \alpha^{\prime} |T|^2}
\label{PotNoWilsonLine}
\ee
where we defined $T=\half \left( T^1+iT^2 \right)$, whereas $T_9$ denotes the tension of a ${\rm D}_9$-brane which is defined, for general $p$, by
\be
T_p=\frac{1}{    (2\pi)^{p}  {\alp}^{\frac{p+1}{2}}   g_s  } 
\ee
where $g_s$ is the string coupling constant.
The stable vacuum is at $T=\infty$, where the vacuum energy vanishes. Since the potential (\ref{PotNoWilsonLine}) is exact, it gives a proof of Sen's conjecture \cite{sen} that the negative energy contribution from the tachyon  precisely cancels the D-brane tension: under tachyon condensation, the D-brane will decay into the closed string vacuum without any D-branes, therefore, excitations are described by closed strings alone. \footnote{In a recent paper \cite{sen3}, the authors have investigated tachyon condensation in the separated $\ddbp$ system including the effects of higher level terms, in superstring field theory.} 

Let us turn now to the case of a spatially dependent tachyon. In this case, by a combination of spacetime and gauge rotations one can bring $T^I$ to the following form:
\be
\sqrt{\alpha^{\prime}} T^I = u^I X^I
\label{LinearTachyonProfile}
\ee
where $u^I$ are constants. When the gauge fields are zero, one can compute the partition function (\ref{BasicZ}) using eqs. (\ref{BulkAction}), (\ref{BdryAction}) and (\ref{LinearTachyonProfile}). The result is \cite{Kraus}
\be
Z = 2 T_9 \, \int d^{10} X_0 \, e^{-2 \pi \alpha^{\prime} T \bar{T}} \prod_{I=1}^{2} F(\pi \alpha^{\prime 2} (\partial_I  T^I )^{2})
\ee
where
\be
F(x) = \frac{4^x x \Gamma(x)^2}{2 \Gamma(2x)}
\ee
The partition function allows us to have an expression for the action of the tachyon at all orders in derivatives. However, there is an ambiguity in the expansion, because any term with at least two derivatives acting on $T$ can be added. At quadratic order the result (for the case of coincident $\ddbn$) is unambiguous
\be
S \approx 2 T_9 \, \int d^{10} x\, e^{-2\pi \alpha^{\prime} T \bar{T}} \left[ 1+ 8 \pi \alpha^{\prime 2 } \, \textrm{ln}(2) \,  \partial^{\mu} \bar{T} \partial_{\mu} T + \ldots \right]
\label{ActionLinearT}
\ee
where the expansion
\be
F(x) = 1 +2 \textrm{ln}(2)\,x + {\cal O}(x^2),\, \quad x\rightarrow 0
\ee
has been used.

Now, consider the case where one of the spatial directions, $y$, is wrapped on a circle of radius $\tilde{R} \leq \sqrt{\alp} $
and that we have a constant Wilson line $A$ wrapping the compact direction on say the ${\rm D}_9$ brane. The gauge field strength in (\ref{BdryAction}) vanishes and the only dependence on the gauge field comes from the covariant derivative. We can lift the above expression to include the covariant derivative by simply changing the argument of the function $F$.\\ 
Applying a T-duality transformation along $y$, the gauge field is mapped to the Higgs field which measures the distance $d$ between 
a $\ddbe $ pair, separated along the dual coordinate $\tilde{y}$ with $d \sim |A|$.

 Adopting the normalization of the tachyon field used in \cite{Hashimoto}, the action (\ref{ActionLinearT}) becomes 
\be
S=2 T_9 \int d^{9} x \, dy \,  e^{-|T|^2} \left[ 1 + 2 \alpha^{\prime} |\partial_{\mu} T|^2 + 2 \alpha^{\prime} A^2 |T|^2 \right]
\label{EffAction}
\ee
The potential  term is
\be
V_0(T) =  2 T_9\, e^{-|T|^2} \left[1+2 \alpha^{\prime} A^2 |T|^2  \right] 
\label{V0}
\ee
The extrema are given by
\be
\frac{ \partial  V_0(T)} {\partial |T|   } = 2 T_9 \, |T| e^{-|T|^2} \left(  4 A^2    \alpha^{\prime} - 2  \left(2 A^2 \alpha^{\prime}  |T|^2+1\right) \right) =0
\ee
i.e.,
\be
|T|=0, \quad |T|=+\infty,\quad \textrm{and} \quad |T| =  \frac{\sqrt{2 A^2 \alpha^{\prime} -1}}{\sqrt{2 \alpha^{\prime} } A}
\ee
To study the nature of these extrema, we need to compute the second derivative: around $|T|=0$ we have
\be
\frac{ \partial^2  V_0(T)} {\partial^2 |T|   }|_{|T|=0}  = m^2 = 4 T_9\,  \left(  2 \alpha^{\prime}  A^2 - 1 \right)
\ee
\begin{figure}
\centering
\includegraphics[width=0.45\textwidth]{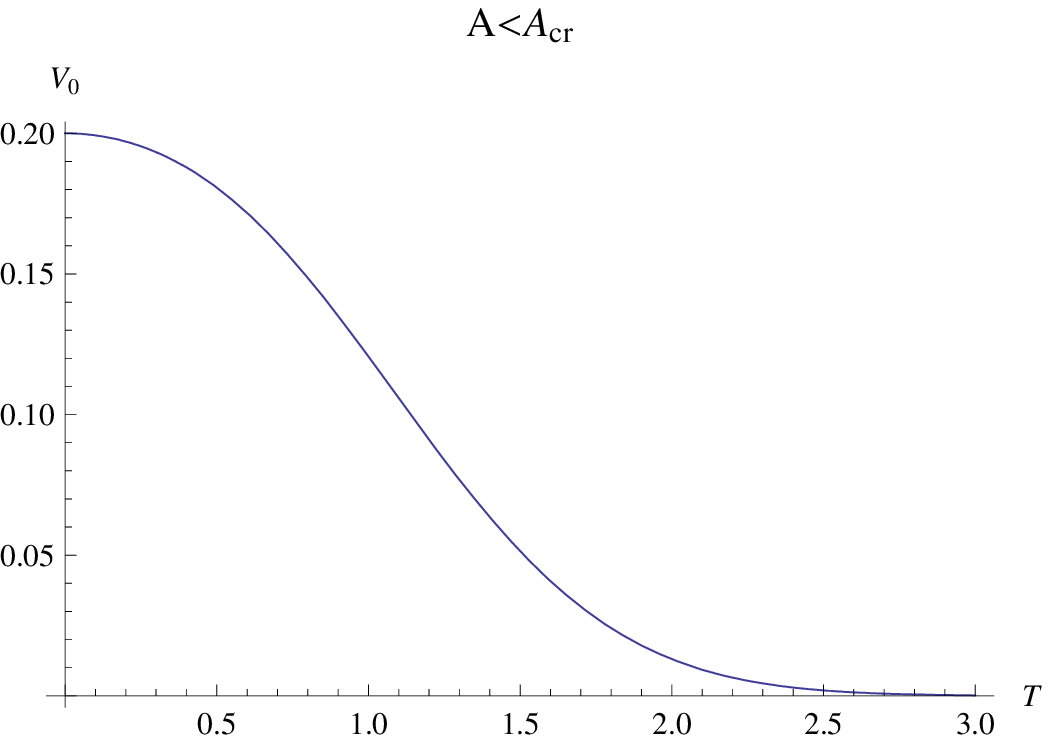} \includegraphics[width=0.45\textwidth]{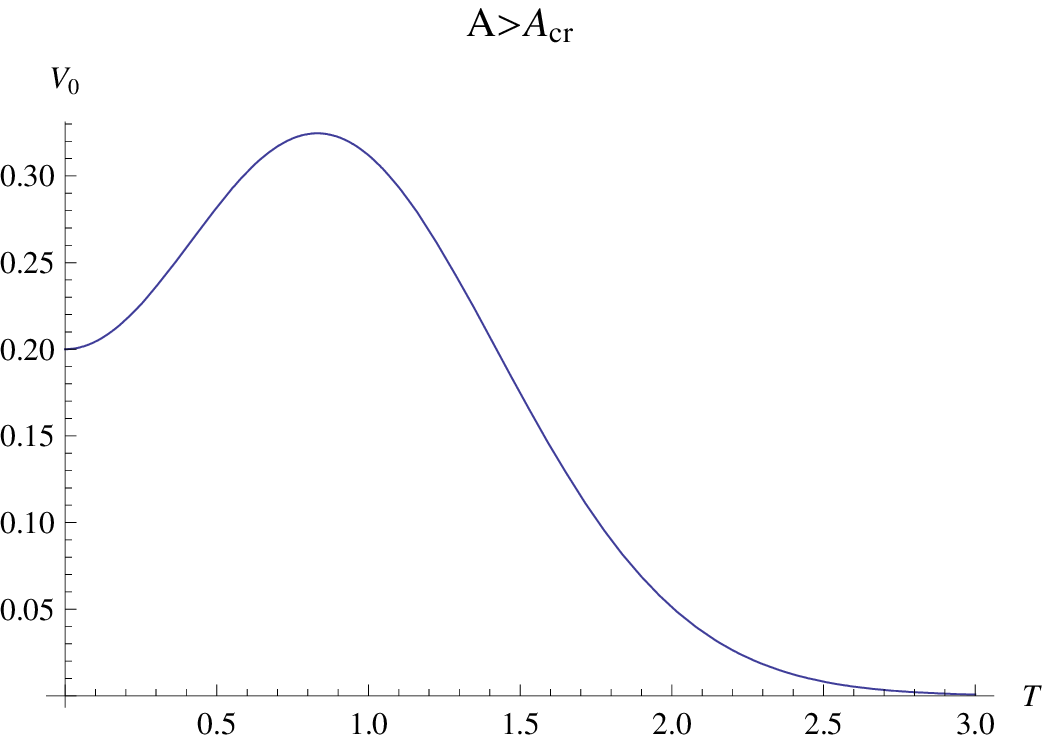} 
\caption{Left: Tachyon potential for $0 \le A<A_{cr}$. Right: $A>\frac{1}{\sqrt{2 \alpha^{\prime}}}$ . In all the plots $T_9=0.1$, $\alpha^{\prime}=1/2$. }
\label{PotZeroTemp}
\end{figure}
Therefore, we see that this potential has a minimum at $|T|=0$ if $A> \sqrt{ \frac{1}{2\alpha^{\prime}} }$ or it has a true tachyonic instability if $A< \sqrt{ \frac{1}{2\alpha^{\prime}} }$.  Figure \ref{PotZeroTemp} shows  the different cases. \\
This behavior has a clear physical interpretation: recall that our model is equivalent to the case of a parallel $\ddbe$ pair separated by a distance $d$. If the distance $d$ is large enough, then the tachyon mode between the two should go away, since the tachyon field comes from the open string suspended between the two branes and thus that string acquires a mass lift when two branes are distant. \\
Notice also that in order to get a canonical kinetic term in the BSFT action we must perform the following redefinition of the tachyon field: $T=T(\phi)$ with
\be
\phi= \sqrt{8 \alpha^{\prime} T_9} \int_0^{|T|} ds \, e^{-s^2/2} 
\ee
With this redefinition, the action (\ref{EffAction}) becomes
\be
S = \int\, d^{9}x \, dy \, \left( \half (\partial \phi)^2  + V_0(T(\phi)) \right)
\ee
and the tachyon vacuum at infinity is placed at a finite value of the new field $\phi$. Indeed, the two local minima are
\be
\phi_0=0\,, \quad \phi_1 = \sqrt{4 \pi  \alpha^{\prime} T_9} .
\ee
This redefinition allows us to compute the mass of the tachyon: in the presence of a Wilson line $A$ it is given by
\be
M^2= \frac{ {\partial ^2 V(\phi)}} {{\partial \phi^2 }} = \frac{1}{\alpha^{\prime}} \left[  \frac{|T|^2-1}{2}  + \alpha^{\prime}  A^2 \left(  |T|^4 - 4 |T|^2 +1\right)  \right]
\label{MassWithA}
\ee
whereas if $A=0$ we have
\be
M^2_{A=0}=  \frac{1}{\alpha^{\prime}}  \frac{|T|^2-1}{2}  
\label{MassWithoutA}
\ee
Notice that the same results were found in \cite{Garousi} but with different methods. \\
Henceforth, we will consider only the real part of the tachyon field: this is consistent with the tachyon equations of motion and it is also a natural setup since we are not interested in lower dimensional D-brane left after the tachyon condensation which needs complex tachyon configurations.


\section{Free energy of open strings stretching between a $\ddbp$ pair}
\setcounter{equation}{0}
\label{secThree}
Before we discuss the free energy of strings stretched between a  $\ddbp$ pair, 
let us first comment on the issue, raised by Hotta in \cite{Hotta1}, concerning the microcanonical ensemble vs canonical ensemble framework for the computation of the tachyon finite temperature potential.
It was shown in \cite{Hotta1} that whilst in principle, the microcanonical picture is more trustworthy as we approach the Hagedorn temperature, in fact for the case of a coincident $\ddbn$ pair, the micro and canonical ensembles agree in the nature and existence of the second order phase transition of the tachyon effective potential near the origin. For the case $p < 9$ the case is less clear as the predictions for phase transitions 
do not entirely overlap for the various values of $p$ in the two formalisms. In this case it is better to adopt the microcanonical ensemble as in \cite{Hotta1}, \cite{Hotta2}.

Since we wish to consider the case where we turn on constant Wilson lines around a compact spatial $S^1$ in the 
$\ddbp$ system, we should consider how this affects the predictions made in both frameworks.
In fact in all cases, the additional terms in the free energy (the singular part of which is 
used in \cite{Hotta1}, \cite{Hotta2} to extract the density of states in the microcanonical ensemble)
 coming from the Wilson line $A$ can be computed. As we shall see below, the Wilson line  only appears as an effective shift in the tachyon mass term when one considers the sums over all states contributing to the 1-loop partition. As such one can verify that at least in the $p=9$ case, the canonical and microcanonical 
formalism will agree as regards the nature of the phase transitions in the tachyon effective potential even with $A \neq 0$. 
For $p<9$ and $A \neq 0$ one should again adopt the microcanonical ensemble too. 
It is straightforward to extend the techniques and results in \cite{Hotta1} to include $A$ but for
brevity we will simply use the canonical ensemble in this paper, in which the one-loop part of the tachyon effective potential is given by the free energy of open strings. 

Thus we will primarily focus on the $p=9$ case in this paper and do all computations in the canonical ensemble.
By T-duality this is equivalent to separated $\ddbe$ pair with separation $d \sim A$. It is interesting to see whether the finite temperature could drastically modify the tachyon potential: in the case of zero Wilson line this is motivated by the fact that the tachyon field at $T=0$ can become stable and there is no tachyon condensation \cite{Danielsson}, \cite{Hotta1}. In the presence of a brane separation, we might also be interested in the fate of the metastable minimum at $T=0$ (see Figure \ref{PotZeroTemp}).\\

Temperature corrections to the potential (\ref{V0}) come from the evaluation of the path integral (\ref{BasicZ}) over all connected graphs of strings on the space where the Euclidean time direction is compactified on a circle of radius equal to the inverse of the temperature $\beta$. We will consider the weak coupling approximation in which the strings can be thought as an ideal gas, that is to say, ignoring the interactions of open strings. We take into account only one-loop amplitude considering only zero-genus oriented Riemann surfaces. 

Let us first quickly review the situation for a coincident $\ddbp$ pair (though  following our discussion above we will ultimately focus on the case $p=9$). The effective potential at finite temperature is given by
\be
V_{eff}(T,\beta) = V_0(T) + V_1(T,\beta)
\label{EffPot}
\ee
where $V_1(T,\beta)$ is the one-loop finite temperature potential.\\ Since we work in the canonical ensemble the one-loop part of the effective potential above is related to the free energy  $F(T,\beta)$ of open strings: 
\be
V_{eff}(T,\beta) = V_0(T) + {\cal V}^{-1} F(T,\beta)
\label{EffPot2}
\ee
where ${\cal V}$ is the volume of the system, the $\ddbp$ pair in our case.\\
At this point,  one immediately faces difficulties: in order to compute $V_1(T,\beta)$ or $F$, we need to include quantum corrections to the BSFT. At tree level the BSFT action is essentially given by the partition function on the disk and at one loop one might expect that the first loop correction corresponds to the partition function on a world-sheet of cylinder or annulus topology. However, because the boundary interactions break conformal invariance this result would depend on the choice of the Weyl factor.\\ Nevertheless, there have been several attempts to generalize the BSFT to the one-loop amplitude in the $D_p-\overline{D_p}$ system \cite{OneLoop}. All of them assume that the relation (\ref{BSFTandAction}) is still true at one loop. Then, they construct the partition function at one loop by keeping  fixed the boundary of the disk and the tachyon profile on it  and adding more boundaries and handles to the string world-sheet diagram. In particular, on the annulus one has
\be
S[u] = \int_{\textrm{annulus}} Z[a_{\textrm{fixed}},b,u]
\ee
where $u$ is the coefficient of the linear tachyon profile eq. (\ref{LinearTachyonProfile}), $a_{\textrm{fixed}}$ is the boundary of the disk and $b$ is the inner boundary of the annulus. Similarly, the cylinder amplitude can be computed in the closed string channel using the boundary state formalism. It is well known that the partition functions obtained in the two different schemes agree on-shell thanks to the open-closed string duality. However, the presence of the tachyon takes the theory off-shell and it is not clear, a priori, that the two different schemes yield the same result. In particular, since the boundary interactions  are due to non-primary fields, the use of conformal maps to transform one worldsheet into another one is not helpful because the transformation laws of the fields are unknown. However, it seems that at one-loop at least, the two results are equivalent.\\
\par
In \cite{Andreev}, for example, the partition function on the annulus and cylinder were computed in the presence of a constant tachyon profile.   
To fix the problems coming from the breaking of conformal invariance, they proposed to use a comparison with field theory results \cite{Zwiebach}, \cite{Gerasimov}. Indeed, given the partition function, one can in principle extract the contribution due to the tachyon and fix the background by comparison with the corresponding field theory results computed from the tree level effective action. This leads to equivalent expressions for the partition functions computed for the annulus and cylinder.\\ 
The one-loop amplitude on the cylinder in such background is given by
\bea
Z_1= -\frac{16 \, i \, \pi^4 V_{p}}{(2 \pi \alpha^{\prime})^{\frac{p}{2} }  } & & \int_{0}^{\infty}  \frac{d\tau}{\tau} \,(4 \pi \tau)^{-\frac{p+1}{2} } \, e^{ -2 \pi  T^2 \tau}  \nonumber \\ && \times \left[     \left( \frac{\theta_3(0 | i \tau) } {\theta^{\prime}_1 (0 | i \tau)}   \right)^4  -  \left( \frac{\theta_2(0 | i \tau) } {\theta^{\prime}_1 (0 | i \tau)}   \right)^4  \right]
\label{PartitionFunction}
\eea
where $V_{p}$ is the volume of the $D_p$-brane. \\
We would like to extend this result in order to include a more general background, i.e., a background in which the tachyon has a linear profile (\ref{LinearTachyonProfile}) and to the case where the $\ddbpm$ pair are separated along a compact direction (equivalently turning on a Wilson line on the dual circle wrapped by a $\ddbp $ pair)
An easy way to do this comes from the following observation.\\
In \cite{Hotta1}, it was noted that the one-loop amplitude (\ref{PartitionFunction}) can be obtained by considering the free energy of strings stretching between the coincident $D_p-\overline{D_p}$ pair \footnote{We adopt the following definition for the Hagedorn temperature $\beta_H^2 = 8 \pi^2 \alpha^{\prime}$. }
\bea
F(\beta) &=& -\frac{V_{p}}{(2 \pi \alpha^{\prime})^{\frac{p+1}{2} }  } \int_{0}^{\infty}  \frac{d\tau}{\tau} (4 \pi \tau)^{-\frac{p+1}{2} }  \sum_{M_{NS}^2}  \sum_{r=1}^{\infty} \textrm{exp} \left( -2 \pi \alpha^{\prime} M_{NS}^2 \tau - \pi \frac{r^2 \beta^2}{\beta_H^2 \tau} \right) \nonumber \\ &+& \frac{V_{p}}{(2 \pi \alpha^{\prime})^{\frac{p+1}{2} }  } \int_{0}^{\infty}  \frac{d\tau}{\tau} (4 \pi \tau)^{-\frac{p+1}{2} }  \sum_{M_{R}^2}  \sum_{r=1}^{\infty} (-1)^r \textrm{exp} \left( -2 \pi \alpha^{\prime} M_{R}^2  \tau - \pi \frac{r^2 \beta^2}{\beta_H^2 \tau} \right)
\label{FE1}
\eea
with the following mass spectrum
\bea
M_{NS}^2 &=& \frac{1}{\alpha^{\prime}} \left( N_B +N_{NS} + \frac{T}{2}^2 -\half  \right) \label{MNS}  \\ 
M_{R}^2 &=& \frac{1}{\alpha^{\prime}} \left( N_B +N_R +  \frac{T}{2}^2 \right) \label{MR}
\eea
where $M_{NS}$ and $M_{R}$ are the masses of the Neveu-Schwarz and Ramond sectors, respectively, whereas $N_B$, $ N_{NS}$ and $N_R$ are the oscillation modes of the bosons, Neveu-Schwarz fermion and Ramond fermions. Notice that the lowest mode of the $NS$ sector (\ref{MNS}) coincides with the mass of the tachyon field (\ref{MassWithoutA}) of the coincident $\ddbp$ pair. 
\par
This suggests a straightforward generalization of eqs. (\ref{MNS}) and (\ref{MR}) to the case of separated $\ddbpm$. 
The only difference with the case described above is that in our model we have a constant Wilson line turned on on a circle of radius close to the string scale. Therefore, in general, we have to include quantized momenta in the direction parallel to the $\ddbp$ system, and winding modes in the direction transverse to it.
As for the presence of the Wilson line, notice that in the T-dual picture, the dependence on the constant Wilson line $A$ in the tachyon mass (\ref{MassWithA}) factorizes out, so  we require that the lowest mode of the $NS$ sector coincides with the tachyon mass. \\
In the general case of $D$ toroidal-compactified directions and $d$ non-compact ones, the mass spectrum is given by \cite{Hotta2}
\bea
M_{NS}^2 &=& \sum_{I=1}^{p-d}  \left( \frac{m_I}{R_I} \right)^2+ \sum_{i=p-d+1}^{D}  \left( \frac{n_i R_i} {\alp} \right)^2+\frac{1}{\alpha^{\prime}} \left( N_B +N_{NS} +  \frac{T}{2}^2 -\half  +M^2_A \right) \label{MNSWithA}  \\ 
M_{R}^2 &=& \sum_{I=1}^{p-d}  \left( \frac{m_I}{R_I} \right)^2+ \sum_{i=p-d+1}^{D}  \left( \frac{n_i R_i} {\alp} \right)^2+ \frac{1}{\alpha^{\prime}} \left( N_B +N_R +  \frac{T}{2}^2  + M^2_A\right) \label{MRWithA}
\eea 
where we have defined 
\be
M^2_A= \alpha^{\prime}  A^2 \left(  T^4 - 4 T^2 +1 \right) 
\ee
Inserting these two expressions into eq. (\ref{FE1}) and expressing the sums in terms of the $\theta$-functions using the conventions of \cite{Hotta1}, the free energy can be written in the following way:
\bea
F(T,\beta)&=&-\frac{16 \pi^4 V_{d}}{(\beta_H)^{d+1} }  \int_{0}^{\infty}  \frac{d\tau}{\tau^{\frac{d+3}{2} } } \, \textrm{exp}^{- \pi \left[ T^2 + 2 M^2_A \right]  \tau}  \prod_{I=1}^{p-d} \theta_3  \left(0|  \frac{2i\alp \tau}{R^2_I} \right)\,  \prod_{i=p-d+1}^{D} \theta_3  \left(0|  \frac{2i R_i^2 \tau}{\alp} \right)      \nonumber \\  & \phantom{1} & \phantom{12345678910111}  \times \left[     \left( \frac{\theta_3(0 | i \tau) } {\theta^{\prime}_1 (0 | i \tau)}   \right)^4 \left( \theta_3(0| \frac{i \beta^2}{\beta^2_H \tau} )-1\right)  \right. \nonumber \\ &\phantom{1} &  \phantom{12345678910111111} \left. -  \left( \frac{\theta_2(0 | i \tau) } {\theta^{\prime}_1 (0 | i \tau)}   \right)^4  \left( \theta_4(0| \frac{i \beta^2}{\beta_H^2 \tau} )-1\right) \right]
\label{FE2}
\eea
where $V_{d}$ is the volume in the non-compact directions parallel to the $\ddbp$ system. \\
This expression for the open string free energy will be our starting point in order to compute the phase transitions in the model under consideration.

\section{Phase transitions at finite temperature}
\setcounter{equation}{0}
\label{secFour}
Given the explicit form of the effective potential eq. (\ref{EffPot2}), it is interesting to see whether temperature corrections could modify the tachyon potential. We expect that at high temperature the system is in a local minimum of the temperature-dependent part of eq. (\ref{EffPot2}). Then, as the temperature decreases, a point will be reached at which a second order phase transition will occur. The critical temperature ${\cal T}_c$ for this to happen, as well as the relevant field space position $T_c$ can be found by solving the following set of equations:
\be
V^{\prime}_{eff} ( {\cal T}_c, T_c ) =0 \quad  \textrm{and} \quad  V^{\prime \prime}_{eff} ( {\cal T}_c, T_c ) =0
\ee
where $V_{eff}$ is given in eq. (\ref{EffPot2}), and the $\prime$ denotes $d/dT_c$.\\
In particular, in the case $A=0$ we expect that temperature corrections should lead to an effective potential in which the location of the minimum has shifted away from infinity. The physical reason for this is that moving towards $T=0$ can be thermodynamically favorable: it costs energy, but it also reduces the mass of the tachyon and therefore increases the entropy of the tachyon gas \cite{Danielsson}, \cite{Hotta1}. We will show explicitly that for temperature near the Hagedorn temperature the minimum will be shifted all the way to $T=0$, in which case the open string vacuum would be stable. \\
In the presence of a separation, it is interesting to see whether as the temperature decreases the system will start rolling towards one or other of the zero temperature minima. \\
\subsection{Low Temperature}
As a warm up calculation and in order to check that our expression for the free energy of open strings, eq. (\ref{FE2}), reproduces known results in the limit of small separation between the $\ddbpm$ pair (equivalently small $A$ in the $\ddbp$ T-dual system), let us study the low temperature approximation of eq. (\ref{FE2}). \\
In \cite{Danielsson}, \cite{Hotta1} it is shown that starting with the minimum of the potential (\ref{PotNoWilsonLine}) at $T=\infty$ and at zero temperature, as the temperature increases the vacuum is shifted from $T = \infty$ to $T=0$. In particular, it is shown that the position of the tachyon minimum, $T_{\textrm{min}}$, moves almost linearly towards $T=0$ as the temperature increases.

In this subsection we will recover this result in the more general background in which a Wilson line is present. \\
In the large $\beta$ limit, we can approximate the free energy (\ref{FE2}) by the large $\tau$ contributions to the integral. In this limit the $\theta$-functions become
\bea
\theta_1^{\prime} (0 | i \tau) &\approx& 2 e^{- \frac{\pi \tau}{4}} \nonumber \\
\theta_2 (0 | i \tau) &\approx& 2 e^{- \frac{\pi \tau}{4}} \nonumber \\
\theta_3 (0 | i \tau) &\approx& 1+ 2 e^{- \pi \tau } \nonumber \\
\theta_4 (0 | i \tau) &\approx& 1- 2 e^{- \pi \tau }
\eea
Using the above expressions, the free energy becomes
\be
F(T,\beta) \approx - \frac{16 \pi^4 V_{d}}{\beta_H^{d+1}} \int_0^{\infty}d\tau\, \tau^{-\frac{d+3}{2} } \textrm{exp} \left[-  \pi \left(T^2  +2 \alpha^{\prime}  A^2 \left(  T^4 - 4 T^2 +1 \right)   -1 \right) \,  \tau - \pi \frac{\beta^2}{\beta_H^2 \tau} \right]
\ee
This integral can be rewritten in terms of the modified K-Bessel function as
\be
F(T,\beta)= -4 V_{d} \left( \frac{   \pi \sqrt{2f(T, A)  -1 } } {\beta_H \beta } \right)^{\frac{d+1}{2}}  \,    K_{\frac{d+1}{2}}   \left( \frac{ 2 \pi  \sqrt{2 f(T,A) -1 }   } {\beta_H} \beta \right)
\label{FreeEnergy0}
\ee
where we defined
\[
f(T,A)=   \frac{T}{2}^2  + \alpha^{\prime}  A^2 \left(  T^4 - 4 T^2 +1 \right)
 \]
In the limit in which both $T$ and $\beta$ are very large the free energy becomes
\be
F(T,\beta) \approx -\frac{ \pi^{\frac{d+1}{2} } V_{d}   }  {\beta_H^{\frac{d}{2}} \beta^{\frac{d}{2}   +1} } (2 f(T,A))^{\frac{d}{2} }   \textrm{exp} \left(-\frac{2 \pi  \beta}{\beta_H} \sqrt{2 f(T,A)}  \right)
\label{FreeEnergy1}
\ee
Inserting this expression in the effective potential (\ref{EffPot2}) and minimizing it wrt $T$ leads to the following condition
\be
T_{\textrm{min}}^2-\frac{2 \pi  \beta}{\beta_H} \sqrt{2 f(T_{\min},A)} =0
\label{CondOnFt}
\ee
In the case $A=0$ we simply have $f(T, A=0 )= \frac{T}{2}^2$ and therefore we get:
\be
T_{\textrm{min}}=\frac{2 \pi  \beta}{\beta_H} 
\ee
If $A \neq 0$, let's assume that its absolute value is $A < \frac{1}{\sqrt{2 \alpha}}$: we are in the regime in which $T=0$ is a maximum of the potential and the tachyon has negative mass near the origin. 
Then, if $T$ is large but $(A^2 \, T^2) \ll 1$, we have
\[
T_{\textrm{min}}^2 -\frac{2 \pi  \beta\, T_{\textrm{min}}}{\beta_H} \sqrt{1+  \, 2\alpha^{\prime} A^2 T_{\textrm{min}}^2} \approx 0
\]
and by expanding the square root
\be
T_{\textrm{min}}^2 -\frac{2 \pi  \beta\, T_{\textrm{min}}}{\beta_H} \left( 1+ \alpha^{\prime} A^2 T_{\textrm{min}}^2 + {\cal O}  ( A^4 \, T_{\textrm{min}}^4 ) \right) \approx 0
\ee
The solution of the previous equation is either $T_{\textrm{min}}=0$, which is not in the $T \gg 1$ approximation, or
\be
T_{\pm\, ,\textrm{min}} = \frac{\beta _H \pm \sqrt{\beta _H^2-16 \pi ^2 \alpha^{\prime} A^2 \beta ^2}}{4 \alpha^{\prime} A^2 \pi  \beta }
\label{TminLowTemp}
\ee
Again here we assume that $\beta$ is large, but $\beta\, A \ll 1$ thus the square root can be expanded in terms of $(\beta^2\, A^2)$
 \bea
T_{-} &=& \frac{2 \pi  \beta}{\beta_H}  + \frac{8 \alpha^{\prime} A^2 \pi ^3 \beta ^3}{\beta^3 _H} +{\cal O}( \alpha^{\prime 2 } A^4\, \frac{\beta^4}{\beta_H^4})  \nonumber \\ T_{+} &=&\frac{\beta _H}{4 \pi \alpha^{\prime}A^2   \beta } -\frac{2 \pi  \beta}{\beta_H}  -\frac{8 A^2 \pi ^3 \beta ^3}{\beta^3 _H} + {\cal O}( \alpha^{\prime 2 } A^4\, \frac{\beta^4}{\beta_H^4})
\eea

$T_{-}$ is the solution we announced at the beginning of this section and it is in agreement with \cite{Danielsson} and \cite{Hotta1}. We see that, as the temperature increases, this minimum shifts almost linearly towards $T=0$. \footnote{Note that the coefficient of the linear term in $\beta$ in eq. (\ref{TminLowTemp}) differs from  \cite{Hotta1} due to the different normalization we adopted in eq. (\ref{FE2}).} \\ What is the meaning of the other solution, namely $T_{+}$? It has an opposite behaviour compared to the previous results: namely the minimum increases as the temperature increases. In fact, we see that this solution violates the approximation we made, namely $A^{2} T^{2} << 1$ and $\beta \,A \ll 1$. For example, say that $A=10^{-7}$, $\beta \approx {\cal O}(10^3)$ then we get $T_{+} \approx 10^{11}$ which gives $A^{2} T^{2} >> 1$. \\

At low temperature, no phase transition occurs regardless of the value of $A$. To see this, expand eq. (\ref{CondOnFt}) around large $T$, this time taking $A={\cal O}(1)$. Then
\be
\frac{ \partial V_{eff} (T,\beta)}{\partial T} =0 \rightarrow T^{\star} = ( \frac{2 } {\alpha^{\prime}})^{1/4}  \frac{              \sqrt{\pi \beta \left(1-4 A^2 \alpha^{\prime}  \right)  } }    {     \sqrt {   A \left( \beta_H -2 \sqrt{2} A \beta \pi  \sqrt{\alpha^{\prime} } \right) }   }
\ee
Computing the second derivative of the potential in this point, we have:
\be
\frac{ \partial^2 V_{eff} (T,\beta)}{\partial T^2}|_{T=T^{\star}} =0 \rightarrow \beta_{cr}=- \frac{\beta_H}{2\sqrt{2 \alpha^{\prime}} \pi A} 
\ee
which is negative and clearly indicates the absence of a second order phase transition at low temperature.\\

Finally, before moving on to consider the high temperature regime, notice that these results can also be found by considering the tachyon field alone, ignoring the contribution of all other open string modes to the free energy of the system \cite{Danielsson}. The reason is that as long as the temperature is low compared to the Hagedorn temperature, the tachyon has the lowest mass and its contribution is dominant. \\ In this setup,  the effective potential, e.g. of  a $\ddbn$ system is given by the sum of the zero temperature tachyon potential and the free energy of the brane-antibrane system at finite temperature ${\cal T} = \beta^{-1}$.
The one loop free energy density for the tachyonic degree of freedom in $9+1$-dimensional space is given by
\be
{\cal F}(T,\beta) = \frac{1}{\beta} \int \frac{d^9 k}{(2 \pi)^9} \textrm{log} \left( 1- e^{-\beta u_k}\right)
\ee
where $u_k=\sqrt{k^2 +\tilde{M}_{NS}^2}$, and $\tilde{M}_{NS}$ is given by eq. (\ref{MNSWithA}) in which the bosonic degrees of freedom $N_B$ and $N_{NS}$ are set to zero. Expanding the logarithm and performing the integration one gets:
\be
{\cal F}(T,\beta) = - \sum_{n=1}^{\infty} (\beta n)^{5} \pi^{-5} 2^{-4} \tilde{M}_{NS}^{5} K_{5}(n \beta \tilde{M}_{NS})
\ee
where $K_{5}(z)$ is the modified Bessel function.
At low temperature (large $\beta$) we keep only the first term in the previous sum, obtaining
\be
{\cal F}(T,\beta) \approx -2 (2 \pi)^{-5} (\tilde{M}_{NS}/\beta)^5 K_5( \beta \tilde{M}_{NS})
\ee
which agrees with the free energy eq. (\ref{FreeEnergy1}) after we expand it in the limit in which both $T$ and $\beta$ are very large.
\subsection{High Temperature}
We will now use the expression for the free energy eq. (\ref{FE2}) in order to investigate the behavior of the model at high temperature, that is to say, at a temperature close to, but below, the Hagedorn temperature. As we discussed earlier in section \ref{secThree}, in this case the canonical ensemble is generally not reliable and we should adopt the microcanonical description in order to compute thermodynamical quantities. But, as we argued there, for the 
case of $\ddbn$ pairs with constant Wilson line, the canonical ensemble agrees with the computations made in the microcanonical ensemble. Thus we will focus our attention on the 
$\ddbn$ system with $A \neq  0$.\\ 

In contrast to the low temperature case discussed before, we now want to expand the integral in  eq. (\ref{FE2}) near $\tau = 0$. To facilitate this, it is convenient to introduce the variable
\[ t= \frac{1}{\tau} \] 
and consider the large $t$ region expansion. \\
Using the modular transformation of $\theta$ functions and extracting the leading term in the large $t$ region near the Hagedorn singularity, we obtain from (\ref{FE2})

\bea
F(T,x) &\approx&  -\frac{\alpha^{\prime\, d-p+\frac{D}{2} } V_{d} } {2^{\frac{D}{2}-2} \beta_H^{d+1}} \left( \frac{\prod_{I=1}^{p-d} R_I^2}{\prod_{i=p-d+1}^D R_i^2} \right) \times  \nn \\ &\phantom{1}& \phantom{123} \int_{\Lambda}^{\infty}dt \, t^{\frac{D+d-9}{2}}  \textrm{exp} \, \left[ -\frac{ \pi  \left(   T^2 +2 M_A^2 \right) } {t} - \pi \left( x^2-1 \right) t \right]  \times  \nn \\ &\phantom{1}& \phantom{12345556} \prod_{I=1}^{p-d} \theta_3  \left(0|  \frac{i R_I^2 t}{ 2 \alp} \right)\,  \prod_{i=p-d+1}^{D} \theta_3  \left(0|   \frac{ i\alp t }{2 R_i^2} \right) 
\label{FEHighT}
\eea
where $\Lambda$ is a cutoff and we have defined $x= \frac{\beta}{\beta_H}$. In (\ref{FEHighT}) we have in mind the case $p=9,\,d=8,\,D=1$.
\par
We are interested in the behaviour of the system at the origin of field space, namely near $T=0$, therefore, we expand the previous expression around this limit and we keep only the lower order terms. As shown in \cite{Hotta2}, the additional contributions from the quantized winding and momenta in (\ref{FEHighT}) may modify the leading order Hagedorn singularity if the compactification radii are 
much bigger than the string scale. In the case where $ p=9, d=8, D=1$ there is only quantized momenta on the circle (since it necessarily lies in a direction parallel to the $D9$ ). If the radius of this circle is close to the string scale then as shown in 
\cite{Hotta2} the Hagedorn singularity is dominant and the expression (\ref{FEHighT}) becomes

\bea
F(T,x) &\approx& -\frac{C V_p}{\beta_H}  \int_{\Lambda}^{\infty} \, dt \, \textrm{exp} \, \left[- \pi  \left(x^2-1 \right) t  -\frac{2 \alpha^{\prime} A^2 \pi }{t} \right] \times  \nonumber \\  & & \phantom{aaaaaaaaa}  \times \left[  t^{\frac{p-9}{2}} + -\pi \left( T^2 +2 \alpha^{\prime} A^2 \left(  T^4-4   T^2 \right) \right)  \, t^{\frac{p-11}{2}} \right] 
\label{FEHighT0}
\eea
where we have replaced $p=d+D$ and defined 
\bea
C&=&\frac{\alpha^{\prime\, d-p+\frac{D}{2}}   } { 2^{\frac{D}{2}-2} \beta_H^{d} \prod_{i=p-d+1}^D R_i} \nn \\ V_p&=&V_d \prod_{I=1}^{p-d} R_I 
\eea

On the other hand the same will be true if we take the radius $\leq  \sqrt{\alpha'} $ and assume that the energy of our 
system is sufficient to excite the quantized momentum modes along the $S^1$.   
For the case $p=9, d=8, D=1$ the latter condition means we may consider small $R_1$ and large $t$ such that $R^2_1\, t$ is still sufficiently large to allow us to approximate $\theta_3  \left(0|   \frac{ i \,R_1^2 t }{2 \alp } \right)$ by unity. 

Under this assumption, we can expand the exponential containing $A$ in the previous expression, as long as $A \approx {\cal O}(1)$.  Keeping only the first two terms we find:

\be
F(T,x) \approx -\frac{C V_p}{\beta_H}  \int_{\Lambda}^{\infty} \, dt \, e^{-\pi (x^2-1)t }\, \left[  t^{\frac{p-9}{2}} - \pi \left(2 \alpha^{\prime} A^2 \left(T^4 - 4   T^2 +1 \right) +  T^2\right)  \, t^{\frac{p-11}{2}} \right] 
\ee
In the case in which $p=9$ this integral can be easily done and the result is
\be
F(T,x) \approx -\frac{C V_p}{\beta_H}  \left[ \frac{1}{\pi  \left(x^2-1\right)}-\pi  \left(2 \alpha'  A^2 \left(T^4-4 T^2+1\right) +T^2\right) \Gamma \left(0,\pi 
   \left(x^2-1\right) \Lambda \right)  \right]
   \label{FinalFreeEnergy}
\ee
We may fix the cutoff scale $\Lambda$ by comparison with the free energy computed in the microcanonical ensemble with $A=0$, \cite{Hotta1}. In particular, if we set $\Lambda=(2 \pi)^{-1}$, the two results agree. We have now all the ingredients to write the effective potential eq. (\ref{EffPot2}) for the  $\ddbn$ pair with constant Wilson line $A$, in order to study the phase transitions. 
By T-duality this is mapped to a separated $\ddbe$ pair with separation $d \sim |A| $.
\\

The critical temperature $\beta_{cr}$ and the value of the tachyon $T_{cr}$ at which the phase transition occurs can be found by finding the solutions of the equation:
\bea
\frac{\partial V_{eff} (T,\beta)} {\partial T} &=& V^{\prime}=  0 \label{fder} \\
\frac{\partial^2 V_{eff} (T,\beta)} {\partial T^2} &=& V^{\prime \prime} =0 \label{sder}
\eea
We have:
\bea
V^{\prime} &=& T_9 \, e^{-T^2} T   \left(8 A^2 \alpha' -4 \left(2 A^2 \alpha'  T^2+1\right)\right)  + \nn \\ && \phantom{1234} \frac{C \pi}{\beta_H}   \left(2 \left(4 T^3-8 T\right) \alpha'  A^2+2 T\right) \Gamma \left(0,\pi  \left(x^2-1\right) \Lambda
   \right)
\label{FDer}
\eea
A clear critical point is $T_{cr}=0$. Substituting this value in $V_{eff}^{\prime \prime}=0$ gives the following condition
\be
\Gamma \left(0,\pi  \left(x_{cr}^2-1\right) \Lambda \right) = \frac{2 \beta_H T_9 \left(2 A^2 \alpha' -1\right)}{C \pi  \left(8 A^2 \alpha' -1\right)} 
\label{GammaT=0}
\ee
This equation is important, because it allows us to compute an approximated expression for the critical temperature at the point $T_{cr}=0$. \footnote{The divergence in the rhs of eq.(\ref{GammaT=0}) coming from the vanishing of the denominator is only apparent since it is due to the truncation to the second order term in the expansion of eq. (\ref{FEHighT0}) around large $t$. The full expression of the free energy (\ref{FEHighT0}) is not divergent for any value of $A$.} 
However, we note that whereas the lhs of eq. (\ref{GammaT=0}) is positive definite, the rhs is positive definite only when $0 \leq A \leq \frac{A_{cr}}{2}$ and $A>A_{cr}$. \emph{ Therefore there is no phase transition at $T=0$ for $ \frac{A_{cr}}{2} \leq A \leq A_{cr}$}.\\

When $0 \leq A \leq \frac{A_{cr}}{2} $ or $A>A_{cr}$ we can expand the gamma function in eq.(\ref{GammaT=0}) near $x=1$ using the fact that
\[
\Gamma(0,t)=-\gamma - \textrm{log}\,t +{\cal O}(t)
\]
and we find
\be
\beta_{cr} \approx \beta _H \left[    1+ \textrm{Exp} \left( - \frac{16}{\pi \, g_s}    \frac{\left(2 A^2 \alpha^{\prime} -1\right)} {\left(8 A^2 \alpha^{\prime} -1\right)}  -\gamma  \right)  \right]
\label{betacr}
\ee
where we have set $\Lambda=1/2 \pi$. For weak coupling, $g_s$ needs to be small, therefore, the argument of the exponential in the previous equation is large and negative which means that the critical temperature is very close the the Hagedorn temperature. (  
The limit of $A \rightarrow 0$ of this expression gives the results that Hotta found in $\cite{Hotta1}$.)\\ 
Let us try now to find other solutions for the system of equations (\ref{fder}) and (\ref{sder}). Isolating the gamma-function from the first equation and substituting it into the second one gives the following condition for the presence of critical points:
\be
\frac{8 e^{-T^2} T^2 T_9 \left(8 \left(T^4-3 T^2+3\right) \alpha^{\prime 2} A^4+2
   \left(3 T^2-4\right) \alpha^{\prime}  A^2+1\right)}{4 \left(T^2-2\right) \alpha^{\prime}  A^2+1}=0
   \label{OtherTcr}
\ee
Except for the point $T_{cr}=0$, \footnote{The point $T=+\infty$ solves the equation (\ref{OtherTcr}) but it is out the range of our approximation, namely, we have expanded the free energy around $T=0$.} other possible solutions to the previous equation are
\be
T^2_{\pm} = \frac{12 \alpha^{\prime 2} A^4-3
   \alpha^{\prime}  A^2 \pm  \alpha^{\prime} \, A^2 \sqrt{-48 \alpha^{\prime 2} A^4-8 \alpha^{\prime}  A^2+1}}{8 A^4 \alpha^{\prime 2} }
\ee
The argument of the square root is positive definite only for $0<A<\frac{1}{2 \sqrt{3 \alpha^{\prime} }}$, but for these values of $A$ one can verify that $T^2_{\pm}$ is negative, giving an imaginary $T$.\\
We conclude then that there is only a second order phase transition at $T=0$.


\subsection{Phase Structure} 

In order to study and understand the phase transitions within a system consisting of a $\ddbn$-pair at high temperature, standard thermal field theory reasoning can be very useful: the minima of the effective potential at high temperature are located around those values of $T$ which minimize the tachyon mass at zero temperature and hence increase the entropy of the tachyon gas.\\
Recall that the mass of the tachyon in the presence of a Wilson line was given in eq. ($\ref{MassWithA}$) which we rewrite here for our convenience in terms of real $T$: 
\be
M^2= \frac{ {\partial ^2 V(\phi)}} {{\partial \phi^2 }} = \frac{1}{\alpha^{\prime}} \left[  \frac{T^2-1}{2}  + \alpha^{\prime}  A^2 \left(  T^4 - 4 T^2 +1\right)  \right]
\label{MassFormula}
\ee
The extrema of $M^2(T)$ above are given by
\be
T_1=0\,
\ee
and
\be
T_2 = \pm \frac{  \sqrt{ 8 \alpha^{\prime} \, A^2 -1 } }  {   2A \sqrt{    \alpha^{\prime}  }  }
\label{T2}
\ee
The second derivative of eq. (\ref{MassFormula}) evaluated at $T_1$ is $(1-8\alpha^{\prime} A^2)/\alpha^{\prime}$ which is positive for $A<\half A_{cr}$. Therefore, for $A<\half A_{cr}$ we expect that at high temperatures, $T=0$ is a minimum of the effective potential. If instead $A>\half A_{cr}$, the point $T=0$ is a local maximum, the minimum being $T_2 \neq 0$. \\

We will now investigate the phase structure of our system in the 3 cases where ${0 \leq A \leq \half A_{cr}}$, 
${ \frac{A_{cr}}{2}  < A < A_{cr}} $ or ${A > A_{cr}}$ respectively.


\subsubsection{$\displaystyle{0 \leq A \leq \half A_{cr}}$} 

In this case, we know that there is a phase transition at $T=0$ which is also a minimum at high temperature.
\begin{figure}[h]
\centering
\includegraphics[width=0.45\textwidth]{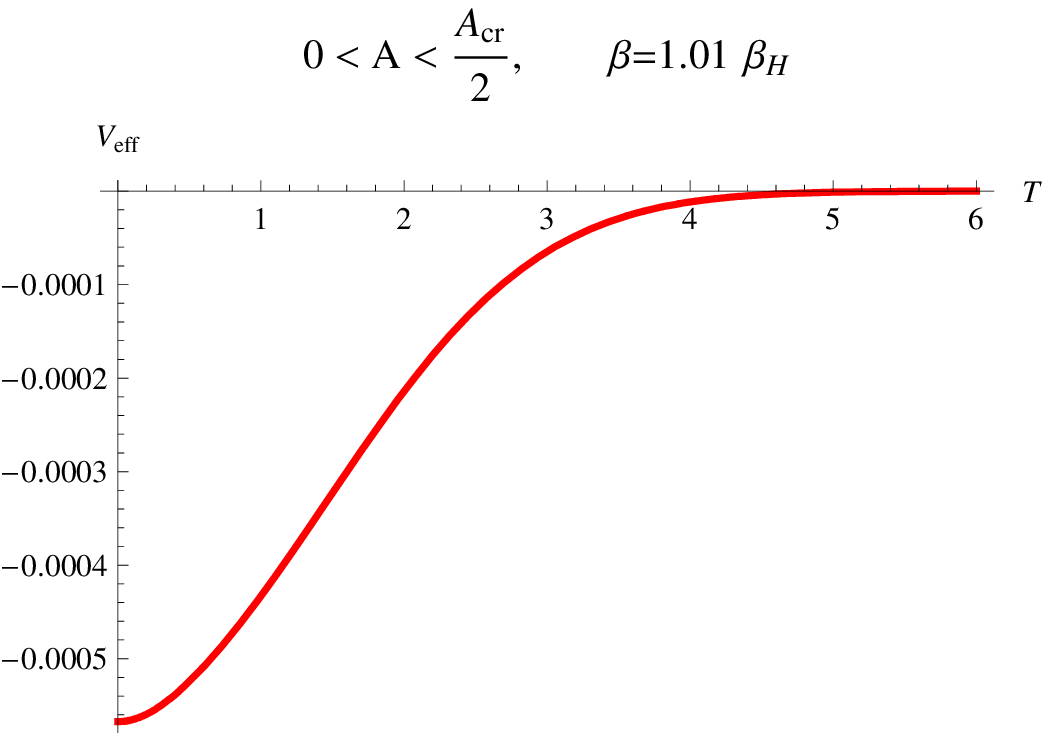} 
\includegraphics[width=0.45\textwidth]{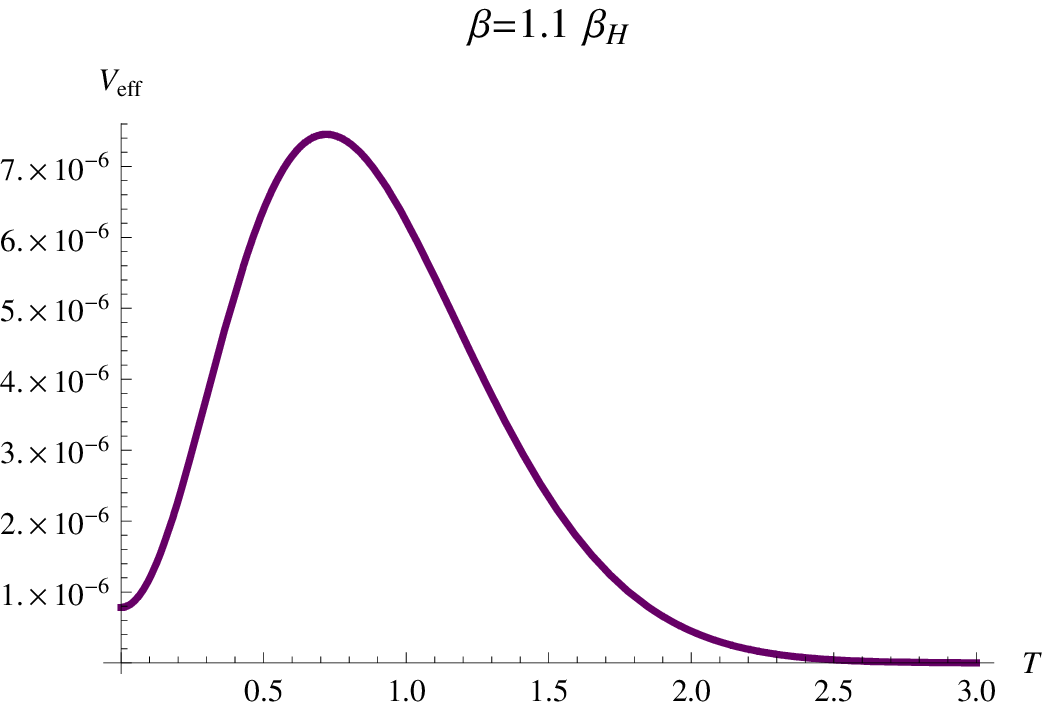}
\includegraphics[width=0.45\textwidth]{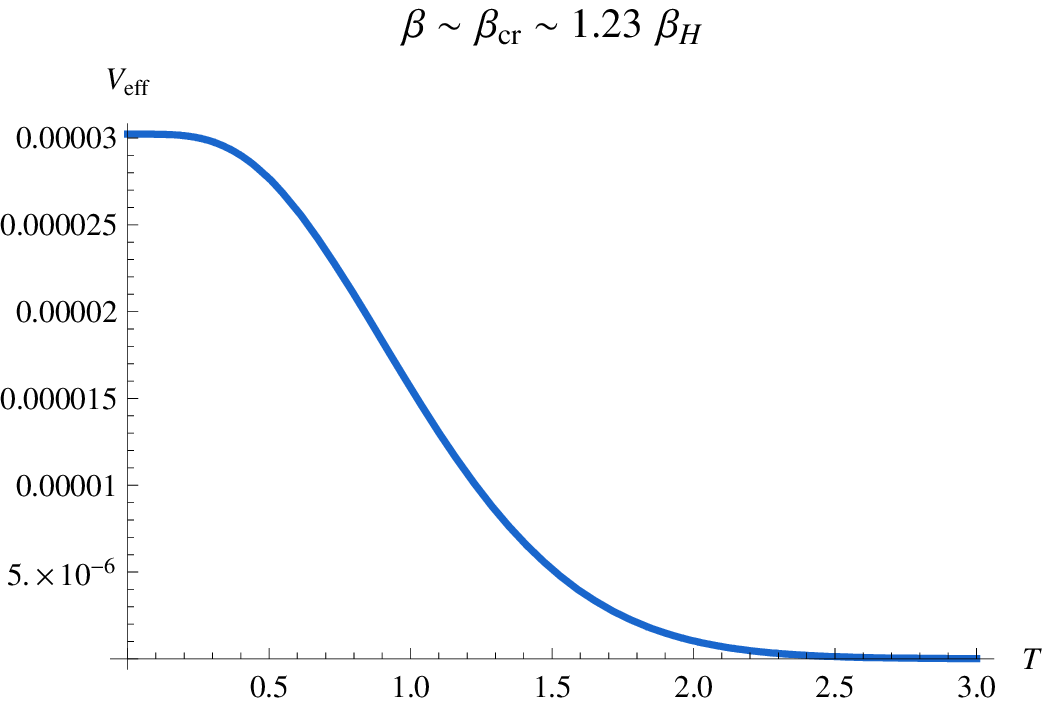}
\includegraphics[width=0.45\textwidth]{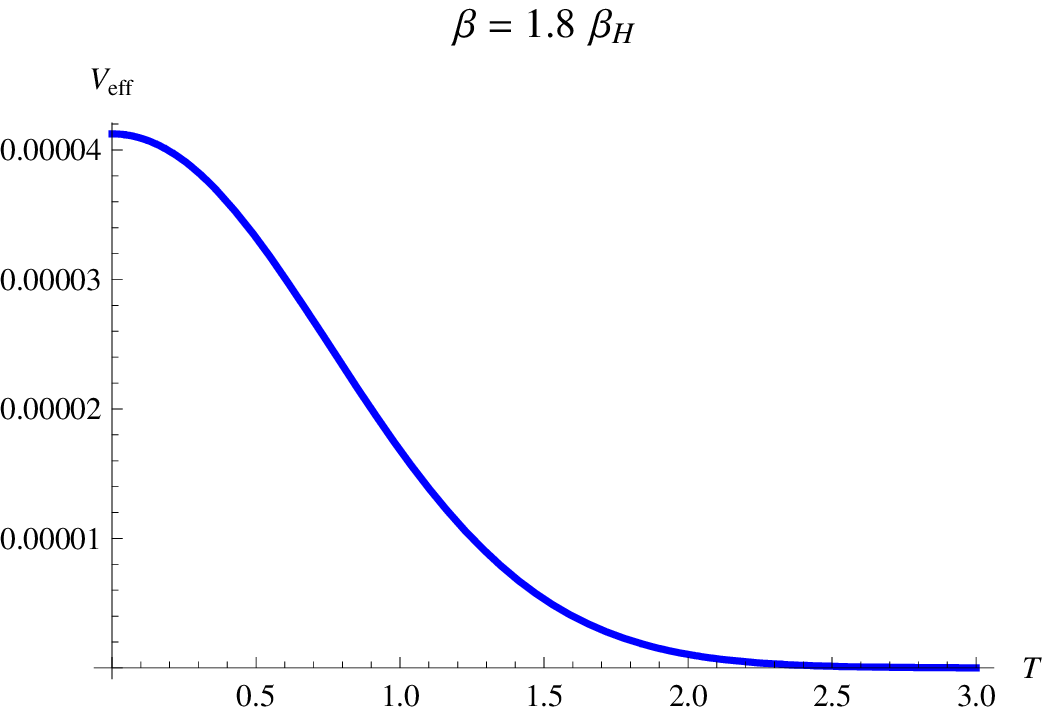}
 \caption{Phase Transition for temperatures close to the Hagedorn temperature and separation, $0 < A < A_{cr}/2 $. 
The plots show the effective potential as derived from eq. (\ref{V0}) and eq. (\ref{FEHighT})
using numerical integration, for various values of the temperature. We chose the values $A=0.3\, A_{cr}$, $g_s=0.1$, $\alpha^{\prime}=1/2$, $\Lambda=1/2\pi$, $p=9$, $d=8$, $D=1$ in these plots.}
\label{ShotsAorigin=0}
\end{figure}

Referring to Figure \ref{ShotsAorigin=0}, we find that:\\
1. \, When the temperature is slightly above to the critical temperature and close to the Hagedorn temperature, i.e. $x \approx 1$, we expect that the system to be in the minimum at $T=0$. Therefore, for these temperatures and for  $0 \leq A \leq \half A_{cr}$ the open string vacuum is stable. \\ We also see in Figure \ref{ShotsAorigin=0} that $T=0$ is actually a global minimum of the effective potential and the latter is negative around this point. To understand why this is the case, consider for example when $A=0$. The zero temperature potential becomes $V_0 |_{ \{T=0,A=0\} }=2 T_9$ and the finite temperature contribution can be obtained from eq. (\ref{FinalFreeEnergy})
\be
F(0, x) |_{A=0} \approx - \frac{C\, V_9}{\pi \, \beta_H \, \left( x^2-1 \right) } \,
\ee
At the critical temperature, this expression can be rewritten using eq. (\ref{betacr}) as
\be
F(0, x_{cr}) \approx -\frac{2 C \, V_9}{\pi \beta_H} \textrm{exp} \left( \frac{16}{\pi \, g_s}-\gamma\right)
\ee
At weak coupling, $g_s$ is small and  consequently the value of the free energy is much larger than the zero temperature piece, resulting in the effective potential becoming negative around $T=0$. \\
2. \, When the temperature is equal to the critical temperature given by (\ref{GammaT=0}) the minimum at $T=0$ becomes flat and is uplifted so that the potential energy becomes positive.  \\
3. \, For temperatures lower than this critical temperature the point $T=0$ is a global maximum and the tachyon field will start rolling towards $T=\infty$ and the system will undergo tachyon condensation.

Moreover, we find that the value of the critical temperature for a second order phase transition at the point $T=0$ is proportional to the value of the Wilson line $A$: the greater the value of $A$, the closer the critical temperature $\beta_{cr}$ is to the Hagedorn temperature. It therefore requires more energy to produce a separated $\ddbe$ than a coincident one.


\subsubsection{$\displaystyle{ \frac{A_{cr}}{2}  < A < A_{cr}} $}
\begin{figure}[h]
\centering
\includegraphics[width=0.47\textwidth]{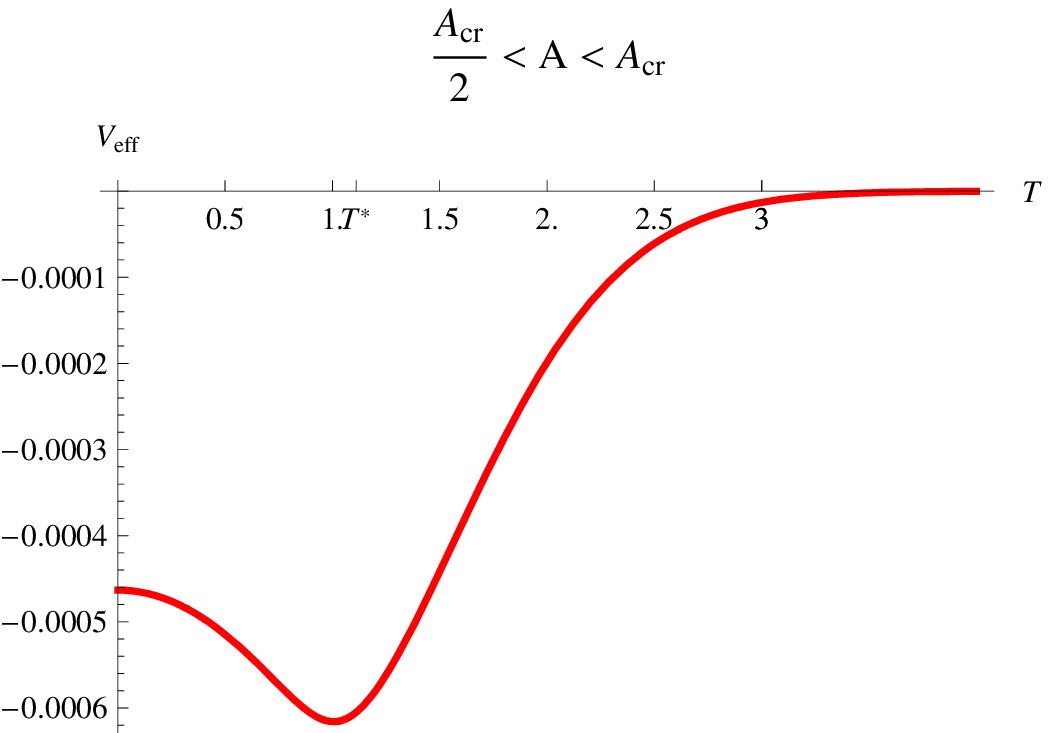} 
\includegraphics[width=0.47\textwidth]{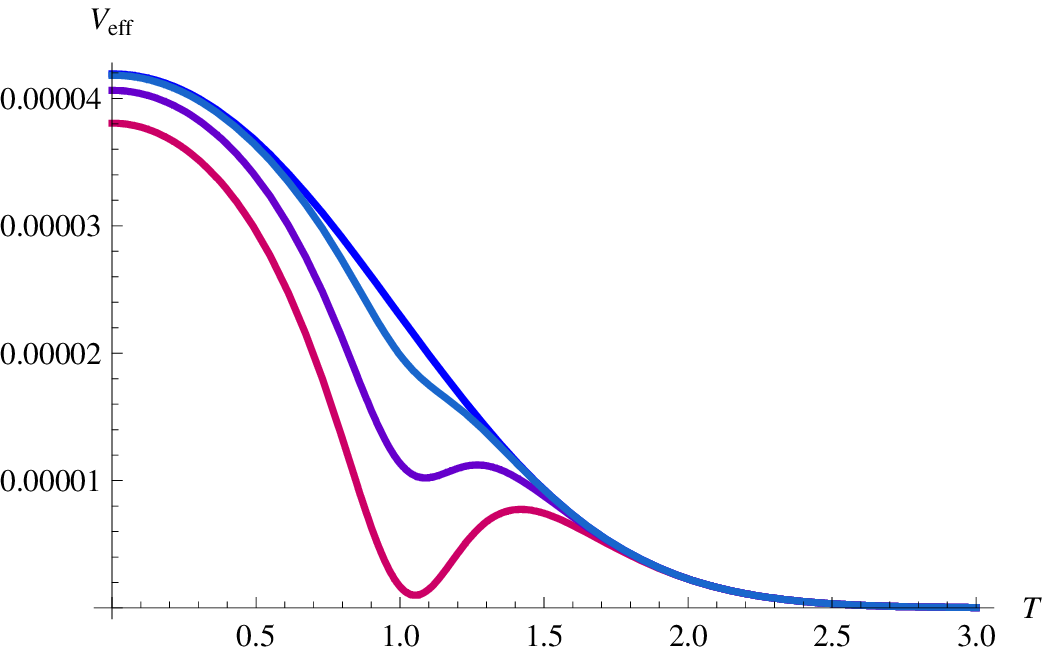}
 \caption{Phase Transition for temperatures close to the Hagedorn temperature and separation, $ \frac{A_{cr}} {2}  < A <  A_{cr} $.
The plots show the effective potential as derived from eq. (\ref{V0}) and eq. (\ref{FEHighT})
using numerical integration, for various values of the temperature and for the choice
$A=0.7\, A_{cr}$, $g_s=0.1$, $\alpha^{\prime}=1/2$, $\Lambda=1/2\pi$, $p=9$, $d=8$, $D=1$. The temperatures are the following: in the left hand plot, $\beta = 1.01 \, \beta_H$; In the right hand plot, $\beta = (1.2,\, 1.3,\, 1.6,\, 2.8) \, \beta_H$  for the sequence of curves displayed from left to right.}
\label{ShotsA1}
\end{figure}

In this case, we know from eq. ({\ref{GammaT=0}}) that there is no phase transition at $T=0$, in fact there is no second order phase transition at all. At high temperatures, thermodynamic reasoning tells us that the system is in a global minimum of the effective potential which is located at $T^{\star} \approx T_2$ given in eq. (\ref{T2}) while the point $T=0$ is a local maximum.
Then, when the temperature decreases this minimum is uplifted and becomes shallower and shallower until it disappears at lower temperatures giving the zero temperature potential as the effective potential. (See Figure \ref{ShotsA1}.)


\subsubsection{     $\displaystyle{A > A_{cr}}$ }

\begin{figure}[h]
\centering
\includegraphics[width=0.47\textwidth]{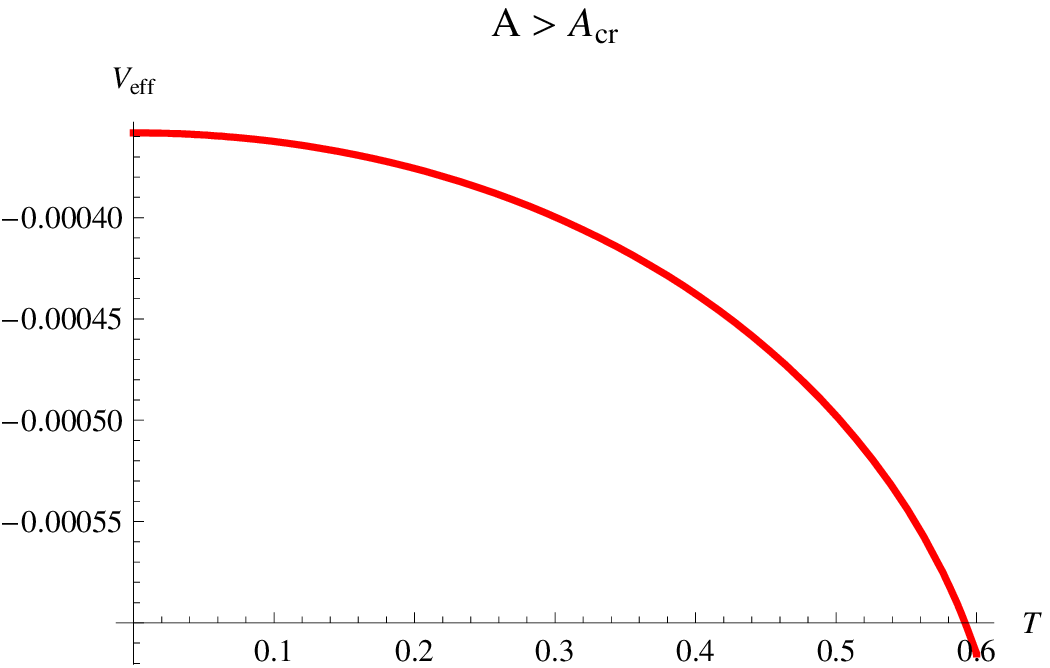} 
\includegraphics[width=0.47\textwidth]{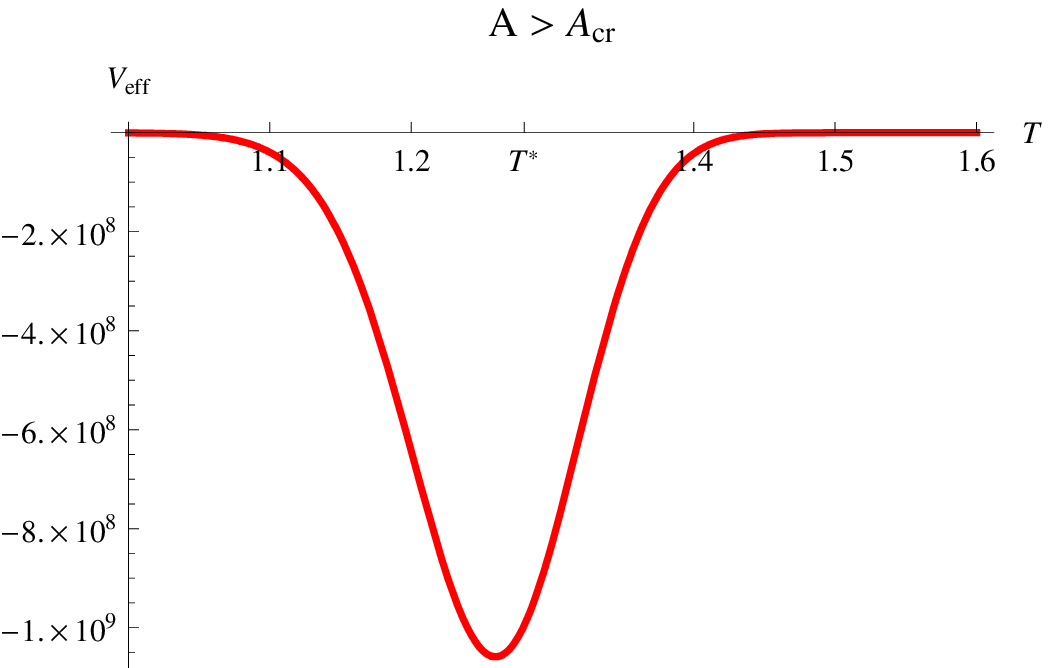}
 \caption{Phase Transition for temperatures close to the Hagedorn temperature and separation, $ A >  A_{cr} $.
The plots show the effective potential as derived from eq. (\ref{V0}) and eq. (\ref{FEHighT})
using numerical integration and  the following parameters choices : $A=1.1\, A_{cr}$, $g_s=0.1$, $\alpha^{\prime}=1/2$, $\Lambda=1/2\pi$, $p=9$, $d=8$, $D=1$. The temperature is the same in both plots: $\beta = 1.01 \, \beta_H$. The left plot is a zoom on the region of the effective potential close to $T=0$ whereas the right one shows the deep minimum at $T=T^{\star}$.}
\label{ShotsAlarge1}
\end{figure}

In this case, referring to Figure \ref{ShotsAlarge1} we find that:\\
1.\, At temperatures very close to the Hagedorn temperature, the system is in a global and deep minimum of the effective potential, say $T^{\star}$, which is not at the origin of the tachyon space, $T=0$. The point $T=0$ is a local maximum at this temperature. (see Figure \ref{ShotsAlarge1}.)\\

\begin{figure}[h]
\centering
\includegraphics[width=0.47\textwidth]{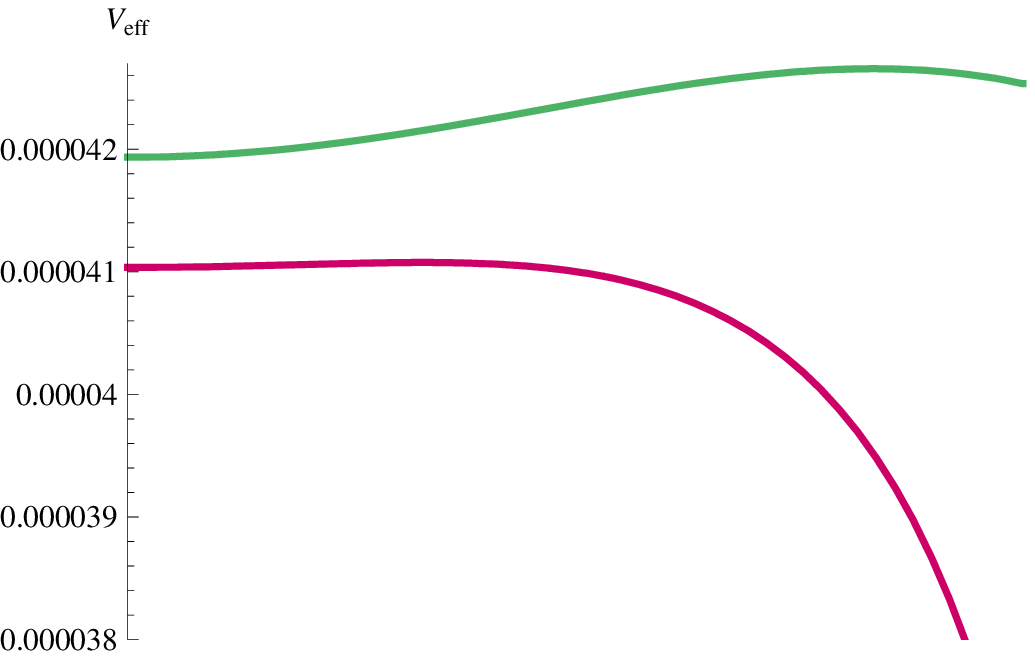} 
\includegraphics[width=0.47\textwidth]{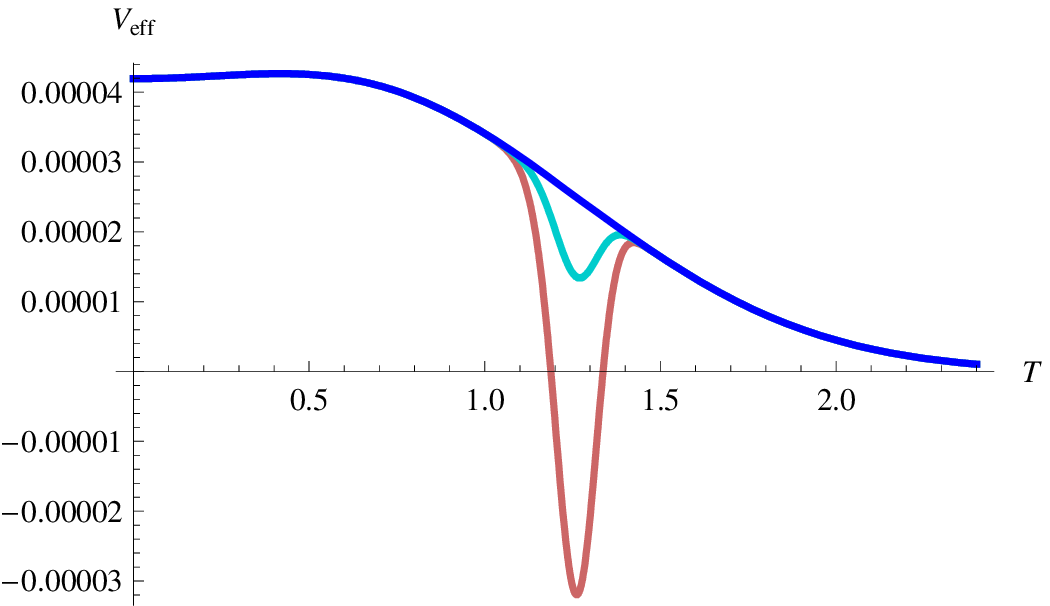}
 \caption{Phase Transition for temperatures close to the Hagedorn temperature and separation, $ A >  A_{cr} $.
The plots show the effective potential as derived from eq. (\ref{V0}) and eq. (\ref{FEHighT})
using numerical integration, with the following parameter choices: $A=1.1\, A_{cr}$, $g_s=0.1$, $\alpha^{\prime}=1/2$, $\Lambda=1/2\pi$, $p=9$, $d=8$, $D=1$. The temperatures are the followings: in the left plots $\beta = (1.199, \, 2.4, \, ) \beta_H$ for the lower and upper curves; in the right hand plot $\beta = (7.8,\, 8.0,\, 8.6) \, \beta_H$  for the three curves starting from the lower.}
\label{ShotsAlarge2}
\end{figure}

2.\,  When the temperature approaches the critical temperature, given by eq. (\ref{GammaT=0}), the point $T=0$ becomes flat and there is a second order phase transition in this point. (see left plot of Figure \ref{ShotsAlarge2}). However, the second minimum continues to exist and it is still deep. \\ 
3. \, For temperatures below this critical temperature, we have two minima $T=0$ and $T=T^{\star}$. (See right plot of Figure \ref{ShotsAlarge2}). As the temperature continues to decrease the minimum $T^{\star}$ becomes shallow and eventually will disappear.  \\
4. \, Eventually, close to zero temperature, the minimum $T^{\star}$ has disappeared and the system will undergo tachyon condensation.

From our prospective, it seems unlikely that at zero temperature the system will be in the open string minimum $T=0$ but rather in the closed string minimum at $T=\infty$. This is because, unless finite temperature tunneling effects happen between the two minima when the separation barrier is short, the system will likely find itself in the minimum $T=T^{\star}$ at high temperature and as the temperature decreases, this minimum will become shallower and the tachyon field will eventually undergo tachyon condensation in the closed string vacuum.



\section{Conclusions}

In this paper we have investigated the phase structure of a $\ddbp$ pair at finite temperature, including a
constant Wilson line $A$ wrapping a spatial circle $S^1 $. By T-dualizing along the $S^1$, this system is mapped to 
a $\ddbpm$ pair where the branes are parallel but separated by a distance $d$ along the dual circle $S^1$ with $d \sim |A|$. Due to the limitations of the canonical ensemble as we take the temperatures close to the Hagerdorn transition, our results are mainly focused on the $p=9 $ case.
The extension to all other values of $p$ can be found by extending the microcanonical ensemble calculations of \cite{Hotta1}, \cite{Hotta2} with the inclusion of a non-vanishing Wilson line $A$.

 We found that the inclusion of $A$ makes the effective potential acquire two minima at finite temperature if $A> A_{cr} $ compared to the situation with $A< A_{cr} $ (which includes the case $A=0$ of coincident $\ddbp$ branes studied in \cite{Danielsson}, \cite{Hotta1}). This raised the question concerning which of the two minima our system is likely to be found. If we consider the case where we are at high temperatures, close to but below the Hagedorn temperature, then there is a single minimum with $T \neq 0 $, indicating the open string vacuum is unstable. As the temperature drops a second order phase transition occurs at the origin $T=0$ where a new minima develops which one can interpret as a meta-stable open string vacuum. However unless there are very special initial conditions it is unlikely that the system can be found in this metastable state but rather the second minimum at $T \neq 0$. The latter coincides with the closed string vacuum $T
\rightarrow \infty $ as the temperature approaches zero.

If instead,  $A \leq A_{cr} $, then we showed that there is a phase transition in $T=0$ only for $0 \leq A \leq \frac{A_{cr}}{2}$. In particular if this condition is satisfied, $T=0$ is a global minimum of the effective potential which is negative at high temperature and the system of a $\ddbn$-pair is stable. Then as the temperature decreases this minimum is uplifted and a second order phase transition occurs. 


Notice that in the dual picture, this implies that the separated $\ddbe$ pairs undergo a phase transition, even in the case that the branes become coincident. This might appear at odds with the results of \cite{Hotta1}, where it was found that no phase transition occurred in a coincident $\ddbp $ pair with $p< 9$. But recall that in the dual system the $\ddbe$ pair have one perpendicular spatial direction compactified on a circle. Thus the branes span all non-compact directions. In \cite{Hotta2},  phase transitions for a $\ddbp$ pair were considered when some spatial dimensions are compactified on a torus. It was shown that a phase transition will occur for a coincident  $\ddbp$ pair even with $p<9$ as long as the branes span all the non-compact directions. Thus our results are consistent with those in \cite{Hotta2}. 

For $\frac{A_{cr}}{2} < A \leq  A_{cr}$ we showed that there is no phase transition near $T=0$. At high temperature the system is in a minimum away from the origin. As the temperature decreases this minimum eventually disappears.

Our analysis in this paper is directed mainly at separated $\ddbe $ pairs because of the limitations of the canonical ensemble for high temperatures. To properly study the case of separated 
$\ddbp $ pairs with $p \leq 7$ (where we assume separation along a single compact direction) will require use of the microcanonical ensemble and extension of the complex temperature techniques used in the case of coincident $\ddbp$ pairs, considered in \cite{Hotta1}. Research in this direction is currently ongoing \cite{calo-thomas}.

Finally, even if the $\ddbe$ system were to find itself in the metastable minimum at $T=0 $ as the temperature decreases, one should then consider the possibility that quantum tunneling effects can lead to the nucleation of closed string vacua $T\neq 0$. In the zero temperature case, \cite{Garousi2} considered the possibility of tachyon tunnelling between the two minima of the effective potential  
when $A> A_{cr} $. It would be interesting to extend this analysis in the case of finite temperature since then the barrier height and width between the two minima becomes a function of temperature so that it is not a priori obvious if tunnelling effects will be suppressed or not.\\ \\

\newpage

{\bf Acknowledgments}

We would like to thank John Ward for very useful comments and discussions.
The work of VC is supported by a Queen Mary studentship.

\end{document}